\documentclass[aps,twocolumn,prb]{revtex4}
\usepackage{graphicx}% Include figure files
\usepackage{dcolumn}% Align table columns on decimal point
\usepackage{bm}% bold math

\begin{document}
\preprint{APS/123-QED}

\title{Unique phase diagram with narrow superconducting dome in EuFe$_2$(As$_{1-x}$P$_x$)$_2$ due to Eu$^{2+}$ local magnetic moments}

\author{Y. Tokiwa}
\author{S.-H. H\"{u}bner}
\author{O. Beck}
\author{H.S. Jeevan}
\author{P. Gegenwart}

\affiliation{I. Physikalisches Institut,
Georg-August-Universit\"{a}t G\"{o}ttingen, 37077 G\"{o}ttingen,
Germany}

\date{\today}% It is always \today, today,

\begin{abstract}
The interplay between superconductivity and Eu$ ^{2+}$ magnetic moments in EuFe$_2$(As$_{1-x}$P$_x$)$_2$ is studied by electrical resistivity measurements under hydrostatic pressure on $x=0.13$ and $x=0.18$ single crystals. We can map hydrostatic pressure to chemical pressure $x$ and show, that superconductivity is confined to a very narrow range $0.18\leq x \leq 0.23$ in the phase diagram, beyond which ferromagnetic (FM) Eu ordering suppresses superconductivity. The change from antiferro- to FM Eu ordering at the latter concentration coincides with a Lifshitz transition and the complete depression of iron magnetic order.  
\end{abstract}

\maketitle

After the discovery of high-$T_c$ superconductivity in iron-based materials,~\cite{Kamihara-JACS08} tremendous amount of research has been performed on its properties.\cite{johnston-AP10,paglione,stewart-RMP11} Following the discovery in LaFeAs(O,F) with $T_c=26$~K,\cite{Kamihara-JACS08} superconductivity was found in many materials with related crystal structures, that commonly possess iron-pnictide or iron-chalcogenide layers. Like in the cuprates and heavy fermion metals superconductivity of the iron-based compounds has an intimate relation to magnetism. The maximal $T_c$ is found in the vicinity of the extrapolated point where spin-density-wave (SDW) order of the Fe 3$d$ magnetic moment is suppressed by pressure or doping.

The AFe$_2$As$_2$ (A=Ba, Sr, Ca or Eu) ("122") systems are prototype iron pnictide materials, since clean, large and homogeneous single crystals are available and various ways of tuning towards SC have been reported.~\cite{johnston-AP10,paglione,stewart-RMP11} EuFe$_2$As$_2$ is unique among them because it caries a local magnetic moment due to the divalent Eu atoms. It exhibits a combined transition of structural and SDW order of Fe magnetic moments at $T_0=190$~K and subsequently Eu 4f moments order below $T_{\rm N}=19$~K into a canted antiferromagnetic (AF) state.\cite{ren-prb08,xiao-prb09,jeevan-prb08,jeevan-prb08_2} This state is characterized by a ferromagnetic (FM) alignment of the moments along the orthorhombic $a$-axis with AF coupling along $c$.~\cite{xiao-prb09} Interestingly, the magnetic susceptibility above $T_{\rm N}$, which is dominated by the fluctuating Eu$^{2+}$ moments, displays a Curie-Weiss law, $\chi=\chi_0+C/(T-\theta)$, with positive Weiss temperature, $\theta\sim20$~K, despite the AF ground state.\cite{ren-prb08,Shuai-NJP11}  Indeed, the AF ground state could easily be switched to a FM state in small in-plane fields of order 1~T.\cite{Shuai-NJP11} These observations suggest that the Eu-system is close to a FM instability. Either the application of hydrostatic pressure to EuFe$_2$As$_2$ or the P-substitution on the As site in EuFe$_2$(As$_{1-x}$P$_x$)$_2$, which induces chemical pressure, suppress the $T_0$ transition and induce superconductivity.\cite{Ren-PRL09,jeevan-prb11,Miclea-prb09,Kurita-prb11,Matsubayashi-prb11,terashima-jpsj09} The superconducting (SC) transition reaches up to 30~K at the optimum pressure and P-substitution around 2.8~GPa and $x\approx 0.2$, respectively.\cite{Miclea-prb09,Kurita-prb11,Matsubayashi-prb11,terashima-jpsj09,jeevan-prb11} The magnetic ordering of Eu moments also changes its character as P is substituted. At low $x$ a canting of the moments along the $c$-axis develops, which grows with increasing $x$, giving rise to a FM hysteresis in magnetization along the $c$-axis, while the system still displays an AF ground state.~\cite{zapf-prb11} At $x>0.23$ the Eu ordering switches to FM.~\cite{jeevan-prb11}

One of the important and controversially discussed issues in EuFe$_2$(As$_{1-x}$P$_x$)$_2$ is the interplay between superconductivity and Eu-FM ordering. In the past, Z. Ren {\it et al.} reported a SC transition at 26\,K, followed by FM ordering of Eu magnetic moments at 20\,K on polycrystalline samples of EuFe$_2$(As$_{0.7}$P$_{0.3}$)$_2$ and discussed a bulk coexistence of both phenomena, which would have important consequences on the SC order parameter.~\cite{Ren-PRL09} Subsequent magnetic Compton scattering experiments on similar polycrystalline material indicated competition between the two phenomena.~\cite{Ahmed-prl10} We have previously reported the phase diagram for single crystalline EuFe$_2$(As$_{1-x}$P$_x$)$_2$.~\cite{jeevan-prb11} In contrast to the report on polycrystals, we found that single crystals with $x\geq 0.26$ are lacking bulk superconductivity. In this study we have carefully investigated the homogeneity of the P content by energy dispersive x-ray (EDX) microprobe analysis, since already a small inhomogeneity in the P content could lead to a seeming coexistence of SC and FM order, due to contributions from different volume fractions. Thus, any small inhomogeneity or deviation between the nominal and actual composition may explain the discrepancy to the experiments on polycrystals. However, three more recent studies on polycrystals also claim a much wider SC region for $0.2\leq x \leq 0.4$ and concluded a bulk coexistence of superconductivity and FM order.~\cite{nowik-jphys11,guanghan-jphys11,munevar} Since this issue may sensitively depend on inhomogeneities and sample quality which could vary with different P-substitutions, we decided to perform detailed hydrostatic pressure experiments on two selected well characterized P-substituted single crystals.

As shown below, we can perfectly map the hydrostatic pressure results to our previously determined phase diagram for single crystalline EuFe$_2$(As$_{1-x}$P$_x$)$_2$. In particular, we verify the peculiar extremely narrow existence range of bulk superconductivity and its suppression at the concentration $x=0.23$ for which Eu magnetic order switches to FM. Additionally, we can relate this transition to a change of the electronic structure due to a Lifshitz transition, that meanwhile has been established by angular-resolved photoemission spectroscopy (ARPES),~\cite{Thirupathaiah-prb11}  as well as thermopower measurements.~\cite{Maiwald-prb12}

\begin{figure}[htb]
\includegraphics[width=\linewidth,keepaspectratio]{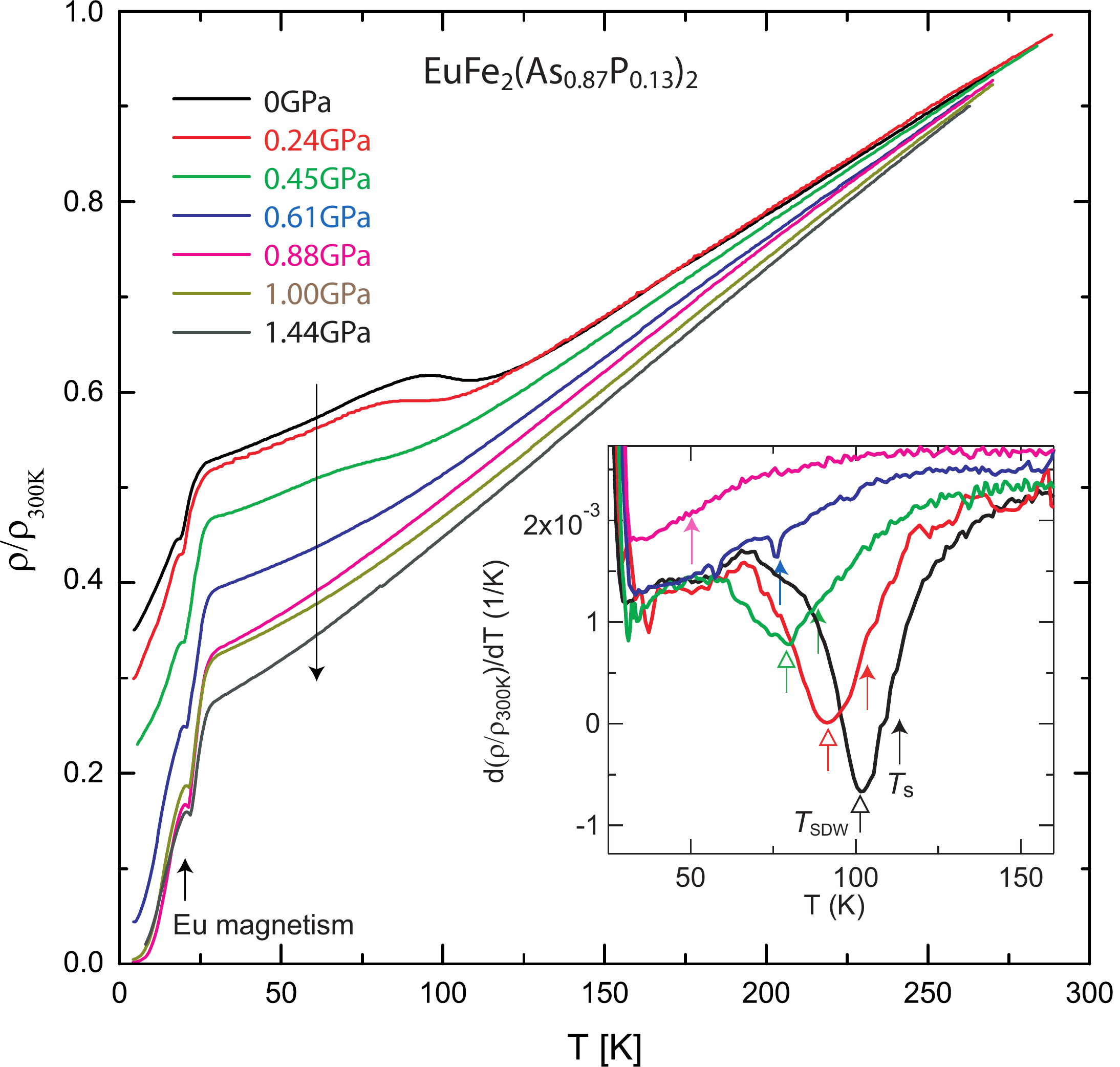}
\caption{(Color online) Electrical resistivity normalized by its value at 300\,K, $\rho/\rho_{300\rm K}$, for EuFe$_2$(As$_{0.87}$P$_{0.13}$)$_2$ under hydrostatic pressure. The inset displays the temperature derivative of the resistivity $d\rho/dT$ vs. $T$. Arrows indicate $T_s$ and $T_{\rm SDW}$ which are determined from the inflection points and minima, respectively.}
\end{figure}

Single crystals of EuFe$_2$(As$_{1-x}$P$_x$)$_2$ were grown by the FeAs self-flux method.~\cite{jeevan-prb11} The homogeneity and actual composition of the two samples with $x=0.13$ and $x=0.18$ was confirmed within $\Delta x=0.01$ error by EDX microprobe analysis on several points of cleaved surfaces. Powder X-ray analysis displays a compression of the unit cell volume, related to the chemical pressure effect of P substitution.~\cite{jeevan-prb11} The temperature dependence of the electrical resistivity under hydrostatic pressure was measured by a standard four probe method with the current flowing in the tetragonal basal plane. The measurements were performed from room temperature down to 4.2\,K and under hydrostatic pressure up to $\sim 1.5$\,GPa by utilizing a CuBe piston-cylinder pressure cell. Daphne oil was used as pressure transmitting medium. The applied pressure was carefully determined by detecting the change of the SC phase transition temperature of a piece of Pb, placed in the pressure cell. From the observation of a pressure independent width of the Pb SC phase transitions, we conclude a good hydrostaticity of the pressure.

\begin{figure}[htb]
\includegraphics[width=\linewidth,keepaspectratio]{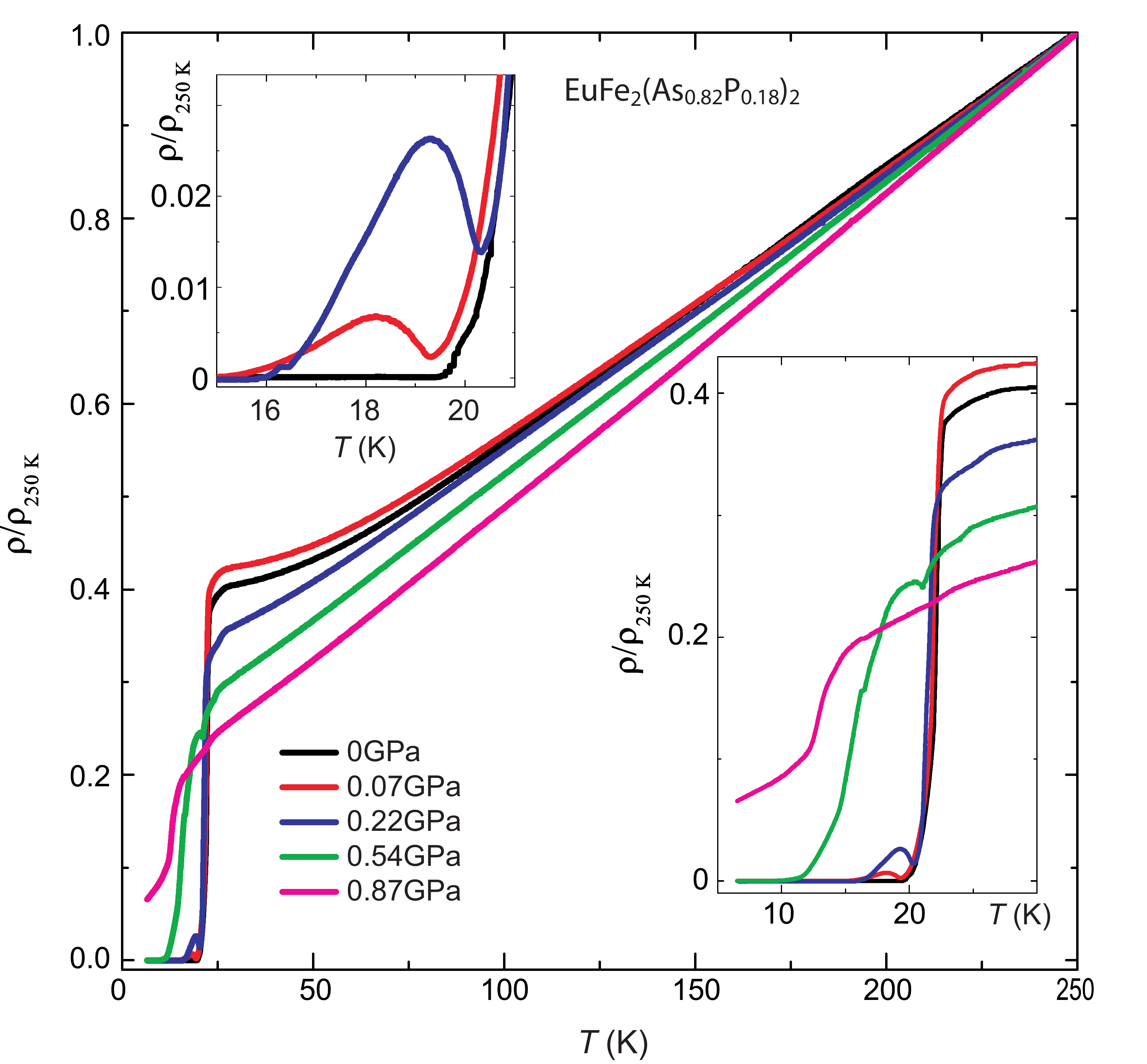}
\caption{(Color online) Electrical resistivity normalized by its value at 250\,K, $\rho/\rho_{250\rm K}$, for EuFe$_2$(As$_{0.82}$P$_{0.18}$)$_2$ under hydrostatic pressure. The two insets enlarge the low temperature region.}
\end{figure}

We first discuss measurements on a $x=0.13$ single crystal displayed in Figure 1. The combined transition at $T_0=190$\,K for the pure compound becomes separated into two transitions of structural ($T_s$) and SDW ($T_{\rm SDW}$) order in P-substituted materials and their separation increases with increasing P-concentration.~\cite{shuai} The temperature derivative of the electrical resistivity is a sensitive probe of the two phase transitions in 122 iron pnictides. As shown e.g. in Ref.~\cite{Fisher}, $T_s$ and $T_{\rm SDW}$ are characterized by an inflection point and minimum in $d\rho/dT$ vs $T$, respectively, compatible with thermodynamic, magnetic and structural experiments. The inset of Figure 1 displays the resistivity derivative of our data, together with arrows at the positions of the inflection points and minima. The signature of the SDW transition could only be resolved up to 0.45~GPa, while a very broadened anomaly related to the structural transition is detectable until 0.88~GPa. The positions of the so-derived transition temperatures in the phase diagram (Fig.~3) are discussed later. Around 30~K, a drop is found in $\rho(T)$, which however does not represent bulk superconductivity, since it is not accompanied by a Mei{\ss}ner signal in the magnetic susceptibility.~\cite{jeevan-prb11} This signature is related to a very small SC volume fraction, likely due to some very small inhomogeneity, which could not be detected within the resolution of X-ray diffraction and EDX. With increasing pressure, the SC signal is getting more pronounced, reaching $\rho=0$ at the highest pressure, indicating an increase of the SC volume fraction. At low temperatures, the resistivity also shows a hump, which is a signature of the Eu magnetic ordering.
%As the pressure is increased, the two transitions at high temperatures ($T_s$ and $T_{\rm SDW}$) are suppressed to lower temperatures and their signatures become weaker. At pressures above 0.61\,GPa, they are not clearly visible any more, although the bending of $\rho(T)$ near 50~K still suggests broadened transitions in this temperature range. Even at the highest pressure of 1.44~GPa, still some curvature in $\rho(T)$ remains below 50~K.

\begin{figure}[tb]
\includegraphics[width=\linewidth,keepaspectratio]{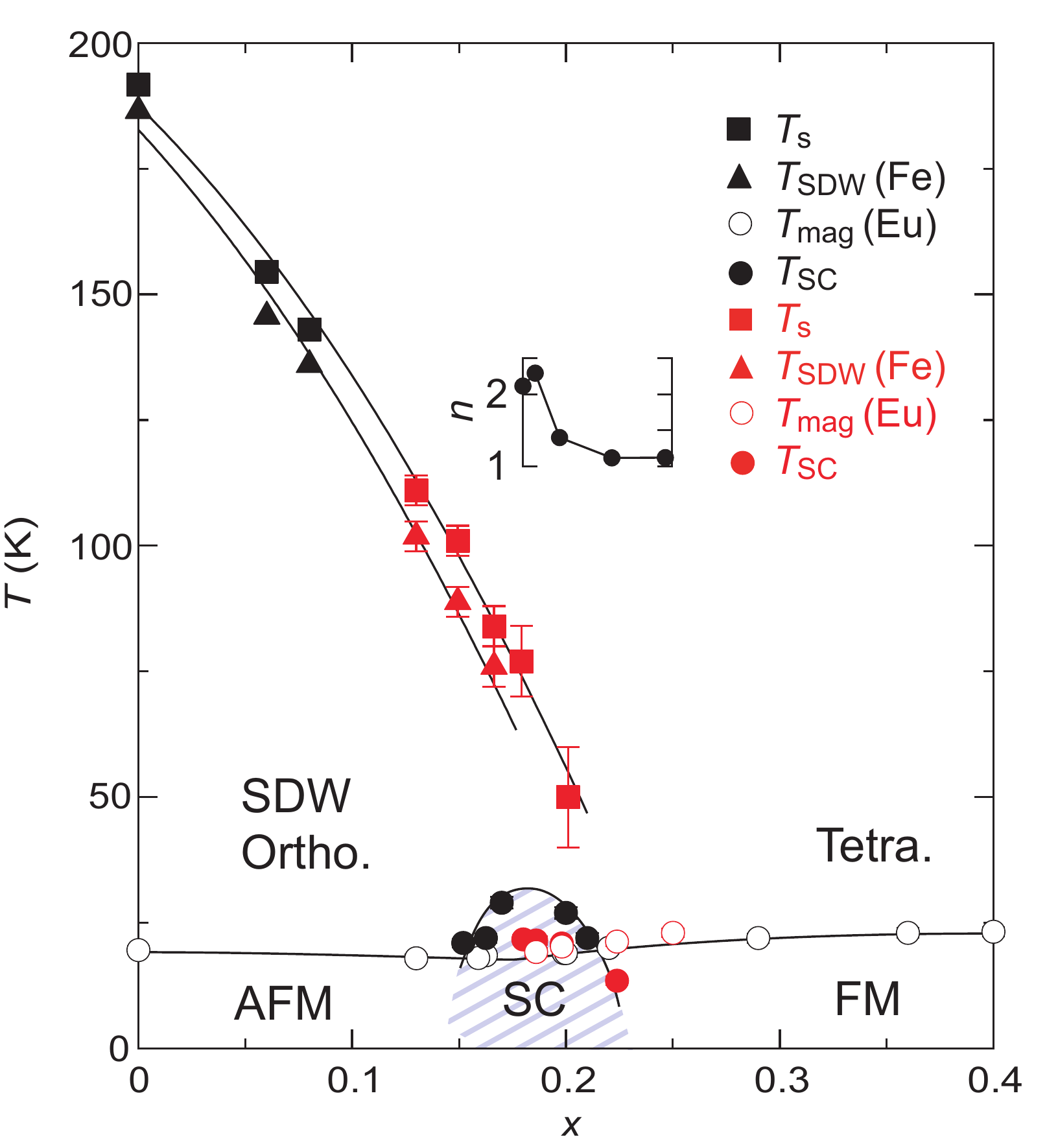}
\caption{(Color online) Phase diagram of EuFe$_2$(As$_{1-x}$P$_{x}$)$_2$, including data points from the previous ambient-pressure work (black symbols) and this study (red symbols). The effective P-concentration $x$ for the data points from this study has been obtained by using the bulk modulus $B=82.9\pm 1.4$~GPa,\cite{Mittal-prb11} and the lattice constants of EuFe$_2$(As$_{1-x}$P$_{x}$)$_2$.\cite{jeevan-prb11} Lines are guides to the eye. The inset shows the exponent $n$ of the temperature dependence of the electrical resistivity $\rho=\rho_0+AT^n$ as a function of the calculated effective P-concentration for EuFe$_2$(As$_{0.82}$P$_{0.18}$)$_2$ under hydrostatic pressure. It shares the horizontal axis for $x$ of the main panel. The exponent $n$ is obtained by fitting the data between 25 and 100\,K.}
\end{figure}

We have performed similar hydrostatic pressure experiments on EuFe$_2$(As$_{0.82}$P$_{0.18}$)$_2$, cf. Figure 2. This sample is a bulk superconductor at ambient pressure, confirmed by Mei{\ss}ner effect and specific heat measurements,~\cite{jeevan-prb11} and displays a sharp and complete resistive SC transition at 22~K, i.e. slightly below the maximal $T_c$ of 30~K found previously for a $x=0.2$ single crystal.~\cite{jeevan-prb11} Upon increasing the hydrostatic pressure, $T_c$ does not increase, as one might have expected from the phase diagram of EuFe$_2$(As$_{1-x}$P$_{x}$)$_2$.~\cite{jeevan-prb11} Rather the SC transition becomes incomplete (cf. the inset), shifts towards lower temperatures and is suppressed at a pressure of 0.87~GPa. The incomplete SC transition also displays a signature at the ordering of Eu$^{2+}$ local moments. In previous hydrostatic pressure experiments on EuFe$_2$As$_2$, similar behavior has been found at pressures slightly below or above the pressure range of bulk superconductivity.~\cite{Miclea-prb09,Kurita-prb11,Matsubayashi-prb11} The data thus indicate, that single crystalline EuFe$_2$(As$_{0.82}$P$_{0.18}$)$_2$ is located just at the border of bulk superconductivity, which disappears at very low pressure.

In order to obtain a quantitative comparison between the pressure and P-substitution, we use the bulk modulus $B=82.9 \pm 1.4$~GPa of BaFe$_2$As$_2$, determined in the orthorhombic state at 33~K (Ref.~\cite{Mittal-prb11}) and the change of the lattice constants in EuFe$_2$(As$_{1-x}$P$_{x}$)$_2$.~\cite{jeevan-prb11} Using these data, 0.61\,GPa corresponds to 5\,\% of P-substitution. Consequently, EuFe$_2$(As$_{0.87}$P$_{0.13}$)$_2$ at a pressure of 0.61\,GPa corresponds to $x=0.18$ at ambient pressure. Indeed both data sets display similar curvature above the SC transition (cf. Figs. 1 and 2), and increasing hydrostatic pressure for both concentrations leads to a more linear temperature dependence of $\rho(T)$. We fitted the resistivity data between 25 and 100 K to a simple power law form $\rho(T)=\rho_0+AT^n$ and followed the evolution of the resistivity exponent $n$ with pressure for the $x=0.18$ single crystal, as shown in the inset of Fig. 3 (note, that the x-axis has been converted to the P content using the above relation). A quasi linear temperature dependence ($n\approx 1$), highlighting non-Fermi liquid behavior, is found for $x=0.18$ at 0.54 and 0.87~GPa, which corresponds to $x=0.224$ and $x=0.25$ at ambient pressure, respectively. This is compatible with existing ambient-pressure data on EuFe$_2$(As$_{1-x}$P$_{x}$)$_2$ single crystals,~\cite{jeevan-prb11,Maiwald-prb12,shuai} which found $n\approx 1.2$ for $x=0.2$, $n\approx 1$ for $x=0.23$, and $n\approx 1.4$ for $x=0.26$. The minimum of $n$ near $x=0.23$ is related to a change of the electronic structure at the concentration, discussed below. In addition, beyond this concentration, the previous ambient-pressure results have proven the suppression of bulk superconductivity (Ref.\cite{jeevan-prb11}), which nicely agrees with the similar behavior for $x=0.18$ under hydrostatic pressure above 0.54~GPa. This indicates an excellent agreement between previous ambient-pressure data on EuFe$_2$(As$_{1-x}$P$_{x}$)$_2$ and the new hydrostatic pressure experiments on the two selected P substitutions. It also confirms, that the narrow existence range of superconductivity in this system is not related to disorder introduced by the P-substitution. Rather it must be related to the change of the electronic and crystal structure with chemical pressure. 
%At ambient pressure for $x$=0.18, it shows a SC transition at 22\,K with zero resistance and no clear signature of magnetism. Under a very small pressure of 0.07\,GPa, a reentrant behavior emerges (see the upper inset in Fig.~2). This resembles the reentrant SC in the rare-earth nickel borocarbides, e.g., HoNi$_2$B$_2$C~\cite{eisaki-prb94}, where SC and magnetism compete. Further increase of pressure enhances the effect of FM on the resistivity and it becomes non-SC at 0.87\,GPa.

For the phase diagram, displayed in Figure 3, we have plotted the previous ambient-pressure phase transition temperatures together with that of our new hydrostatic pressure studies, where the pressure has been converted to an effective P concentration using the above relation. We have determined the SC $T_c$ and the magnetic ordering temperature of the Eu$^{2+}$ moments from the maxima in the temperature dependences of $d\rho/dT$ and $\rho$, respectively. Clearly, superconductivity is suppressed beyond $x=0.23$, in perfect agreement with the previous study.~\cite{jeevan-prb11} The phase transition temperatures $T_{\rm SDW}$ and $T_s$, determined for $x=0.13$ under hydrostatic pressure, are also included to the phase diagram. The extrapolation of both transition temperatures towards absolute zero suggests that their complete suppression happens at $x>0.2$, i.e. beyond the concentration for which $T_c$ is maximal. This is corroborated by the evolution of the resistivity exponent $n(x)$, which has its minimum around $x=0.23$.
Thus, the optimum concentration for superconductivity $x=0.2$ is still {\it within} the orthorhombic phase. For BaFe$_2$(As$_{1-x}$P$_{x}$)$_2$ a much wider SC region in the phase diagram has been found, also under pressure.\cite{Kasahara,Klintberg} Typically for 122 pnictide superconductors, the SC state is found on both sides of the extrapolated QCP where the SDW ordering and structural distortion vanish.\cite{paglione} In EuFe$_2$(As$_{1-x}$P$_{x}$)$_2$, by contrast, superconductivity becomes suppressed at the same concentration $x=0.23$, where resistivity and TEP indicate the most drastic non-Fermi liquid behaviors.~\cite{Maiwald-prb12,shuai} 
%and (within an uncertainty $\Delta x$ of about 0.01) where $T_{\rm SDW}$ and $T_0$ extrapolate to zero temperature (cf. Fig. 3).

The disappearance of iron SDW order is likely caused by a change of the Fermi surface nesting properties detected by ARPES.\cite{Thirupathaiah-prb11} A non-rigid band shift evolution of the Fermi surface with chemical pressure has been found. Importantly, the inner hole-like Fermi surface near the $\Gamma$ point shrinks to zero at $x\approx 0.23$.~\cite{Thirupathaiah-prb11} Indeed, TEP has found indication for a Lifshitz transition near this concentration, from a non-monotonic evolution of $S(x)$ at constant temperature.~\cite{Maiwald-prb12} Such a change of the electronic configuration may also influence the Eu$^{2+}$ magnetic ordering, since density-functional-theory-based calculations have found almost similar ground state energies for the AF and FM configurations using the room-temperature lattice constants in this concentration range.\cite{jeevan-prb11} Since the Ruderman-Kittel-Kasuya-Yosida (RKKY) exchange coupling between the Eu$^{2+}$ local moments is oscillatory with the distance, starting with a FM coupling at low distances, the general trend towards ferromagnetism under chemical pressure is expected. Interestingly, by using a minimal multiband model, it has recently been shown that the Fermi surface nesting has strong influence on the RKKY interaction.\cite{Akbari} Within the SDW phase, the gaping of the Fermi surface induces an anisotropy in the RKKY interaction and modifies its strength and oscillatory behavior with respect to that in the paramagnetic regime. Thus, the suppression of $T_{\rm SDW}$ and $T_s$ near $x=0.23$ is expected to strongly influence the RKKY exchange coupling between the Eu moments and likely the coincidence with the change from AF to FM order is not accidental. Since superconductivity is suppressed by ferromagnetism, this also explains the disappearance of superconductivity at the same concentration.

To conclude, we have compared the previously determined phase diagram for
EuFe$_2$(As$_{1-x}$P$_{x}$)$_2$ with hydrostatic pressure experiments on two selected $x=0.13$ and $x=0.18$ single crystals. Using the reported value of the bulk modulus and the measured change of the lattice constants with $x$, we could quantitatively map the hydrostatic to the chemical pressure in this system. Our pressure experiments confirm the extremely narrow SC dome in this system, which is very different to the respective phase diagram for BaFe$_2$(As$_{1-x}$P$_{x}$)$_2$,\cite{Kasahara,Klintberg} where superconductivity extends over large phase space regions. Our results are in clear contrast to previous reports on the coexistence of superconductivity with FM Eu ordering.\cite{Ren-PRL09,nowik-jphys11,guanghan-jphys11}
%but consistent with a recent study of $\mu$SR and magnetization, which reports a very narrow pressure-induced SC dome of EuFe$_2$(As$_{0.88}$P$_{0.12}$)$_2$, ranging from 0.4 to 0.5\,GPa, well below the critical pressure for the complete suppression of SDW order, $\sim$1.0\,GPa.~\cite{Guguchia-arxiv12}
The evolution of the electrical resistivity exponent $n(x)$ displays a minimum at $x=0.23$.
%, i.e. where $T_{\rm SDW}$ and $T_0$ extrapolate to zero temperature.
At this concentration, ARPES and TEP have detected a Lifshitz transition.\cite{Thirupathaiah-prb11,Maiwald-prb12} The change of the electronic structure together with the structural change most likely modify the Eu-RKKY interaction such that the Eu magnetism switches from AF to FM ordering. Since FM order is incompatible with superconductivity, this explains the peculiar phase diagram of EuFe$_2$(As$_{1-x}$P$_{x}$)$_2$, where a SC phase only exists in a very narrow regime.
% which is located still inside the iron SDW phase.

Collaboration and valuable discussions with M. Dressel, J. Fink, C. Geibel, S. Jiang, D. Kasinathan, J. Maiwald, M. Nicklas, H. Rosner, S. Thirupathaiah, K. Winzer, D. Wu, and S. Zapf are gratefully acknowledged. This work has been supported by the DFG priority program SPP 1458.

\end{document}